\documentclass[oldversion]{aa}
\usepackage{epsfig}
\usepackage{natbib}
\usepackage{lscape}
\usepackage{amssymb}
\usepackage{epstopdf}
\usepackage{longtable}
\usepackage{color}
\usepackage{fancyheadings}
\bibpunct{(}{)}{;}{a}{}{,}

\begin{document}
\title{Probing the effect of gravitational microlensing on the measurements of the Rossiter-McLaughlin effect}

\author{ M. Oshagh\inst{1,2}  \and G. Bou{\'e}\inst{3} \and P. Figueira\inst{1} \and N. C. Santos\inst{1,2} \and N. Haghighipour\inst{4,5}}

\institute{
Centro de Astrof{\'\i}sica, Universidade do Porto, Rua das Estrelas, 4150-762 Porto,
Portugal \\
email: {\tt moshagh@astro.up.pt}
\and
Departamento de F{\'i}sica e Astronomia, Faculdade de Ci{\^e}ncias, Universidade do
Porto,Rua do Campo Alegre, 4169-007 Porto, Portugal
\and
Department of Astronomy and Astrophysics, University of Chicago, 5640 South Ellis Avenue, Chicago, IL, 60637, USA
\and
Institute for Astronomy and NASA Astrobiology Institute, University of Hawaii-Manoa,
2680 Woodlawn Drive, Honolulu, HI 96822,USA
\and
Institute for Astronomy and Astrophysics, Department of Computational Physics, University of Tuebingen,
72076 Tuebingen, Germany
}

\date{Received XXX; accepted XXX}

\abstract {In general, in the studies of transit light-curves and the Rossiter-McLaughlin (RM),
the contribution of the planet's gravitational microlensing is neglected. Theoretical studies, have, however shown that the planet's
microlensing can affect the transit light-curve and in some extreme cases cause the transit depth to vanish. In this letter, we present
the results of our quantitative analysis of microlening on the RM effect. Results indicate that for massive planets in on long
period orbits, the planet's microlensing will have considerable contribution to the star's RV measurements. We present the details of our
study, and discuss our analysis and results.}

\keywords{Planetary system,  stars: brown dwarfs, gravitational microlensing , Techniques: radial velocities, Methods: numerical}

\authorrunning{M. Oshagh et al.}
\titlerunning{The microlensing effect on RM effct}
\maketitle

\section{Introduction}

In a transiting planetary system, aside from decreasing the flux of the light received from the star,
the transiting planet may focus a portion of the light that has been blocked by the planet, through the
microlensing effect \citep{Einstein-36}. In general, in studies of transit light-curves, the contribution of
the planet's gravitational microlensing is neglected. However, it has been shown that in systems with massive
or long period planets, the microlensing amplification of the stellar flux could be significant, and in the era
of high precision photometric observations such as those with the \emph{Kepler}, it could also be detectable
\citep{Sahu-03, Agol-03}. Using an analytical approach, \citet{Sahu-03} have shown that in some extreme
cases, the contribution of the light from the gravitational microlening may be even so large that the transit
light-curve can vanish. This implies that in order to properly analyze the light-curve of a transiting system
and derive its planetary parameters, the microlensing effect has to also be taken into account \citep{Muirhead-13}.

\begin{figure}
  \centering
   \includegraphics[scale=0.35]{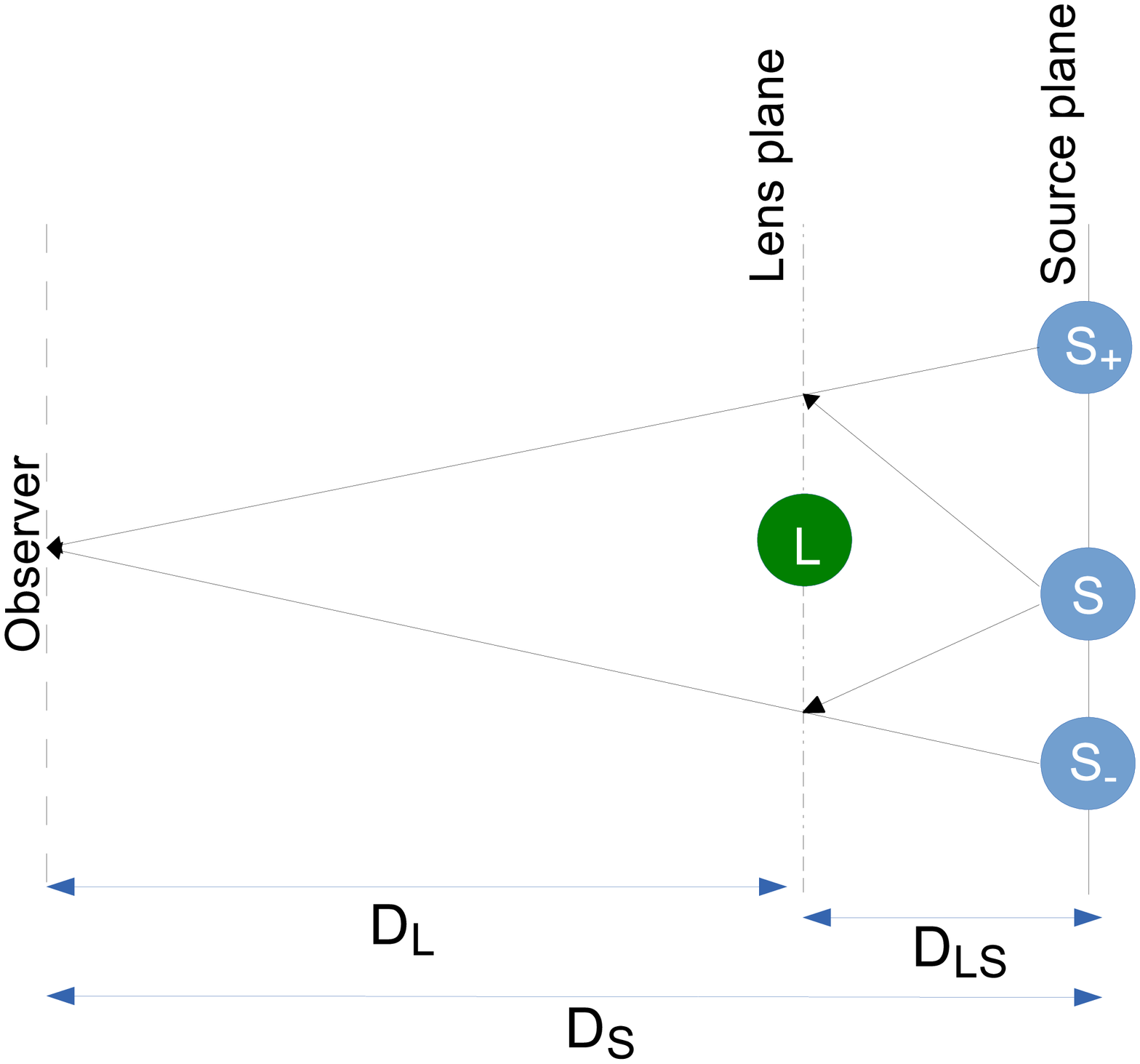}
    \vskip -50pt
   \includegraphics[scale=0.35]{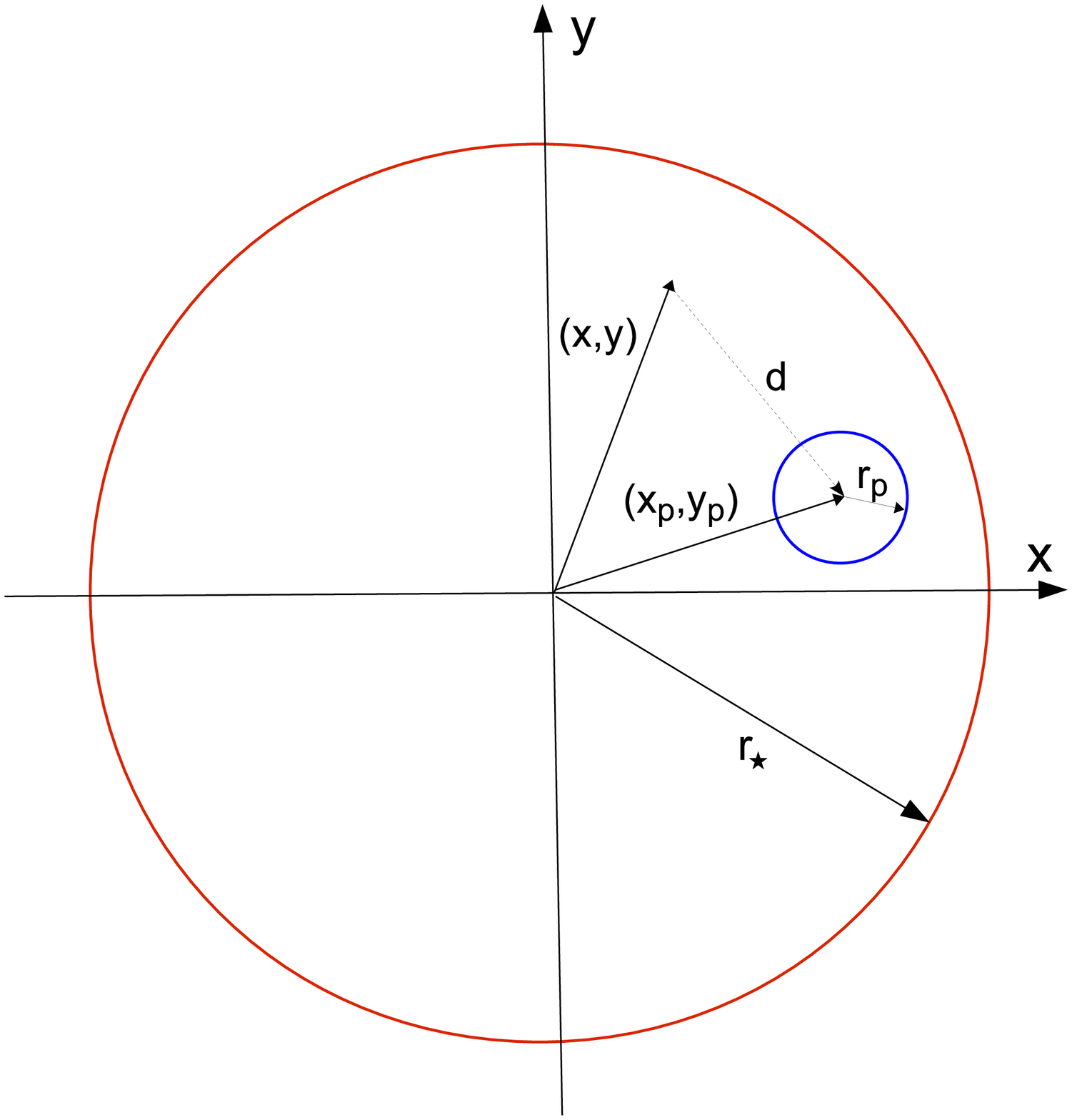}
   \vskip -20pt
  \caption{Top: Schematic configuration of the gravitational microlensing.
Bottom: Schematic view of the coordinate system used in this study. The origin of the (x,y) coordinate
system is at the center of the stellar disk. The quantity $r_{p}$ denotes the planet radius, and $r_{\ast}$ is the radius of the star.}

\end{figure}

One important characteristic of a transiting system that can affected by planet microlensing is the Rossiter-McLaughlin
effect (RM effect).
As a star rotates around its rotational axis, a given point on its surface will be blue-shifted when the star rotates
toward the observer and red-shifted when it rotates away. During a transit, the corresponding
rotational velocity of the portion of the stellar disk that is blocked by the planet is removed from the integration
of the velocity over the entire star creating an anomaly in the Radial Velocity (RV) measurement of the star known as
Rossiter-McLaughlin effect \citep{Rossiter-24, McLaughlin-24}. This effect has been used to constrain the projected
rotation velocity of a star ($v\sin i$), and the angle between the sky-projections of the stellar spin axis and the
planetary orbital plane ($\lambda$) \citep[e.g.,][]{Hebrard-08,Winn-09,Winn-10, Simpson-10, Hirano-11, Albrecht-12}.
 It is important to note the determination of $v\sin i$ and $\lambda$ can be influenced by second order effects
such as the convective blueshift \citep{Shporer-11}, the differential
stellar rotation \citep{Albrecht-12}, and the macro-turbulence \citep{Hirano-11}.

Since the physics and geometry behind the RM effect is due to the transiting planet, they are expected to be affected
by the microlensing effect in the same way. Probing the significance of microlensing on the measurements from the RM effect,
and quantifying its influence
is the main objective of this paper. In section 2, we describe the microlensing effect formalism and introduce
our theoretical model. In section 3, we show the results of our simulations
and discuss the significance of the effect of microlensing and its detectability.
In section 4, we conclude our study and present its future applications.

\section{Model}
We consider a transiting planetary system and assume that the star is an extended source. The surface brightness
of the star is considered to follow a quadratic limb darkening given by

\begin{equation}
I(x,y) = 1.0- u_{1}(1-\mu)-u_{2}(1-\mu)^{2}.
\end{equation}

\noindent
In equation (1), $\mu= \cos\theta=[1-(x^{2}+y^{2})r_{\ast}^{-2}]^{1/2}$, and $u_{1}$ and $u_{2}$ are
the stellar quadratic limb-darkening coefficients. We consider $u_{1}=0.3$ and $u_{2}=0.35$, similar to their values
for the star HD 209458 in the wavelength range of 582 -- 638 nm \citep{Brown-01}.
We also assume that the lens (transiting planet) is an extended object, has zero brightness, and is spherically symmetric.
We set the ratio of the planet's radius to that of the star ($r_{p}/r_{\ast}$) to 0.1, which is similar to the ratio of the
radius of Jupiter or a brown dwarf to that of the Sun. The planet is assumed to be on a circular, edge-on orbit. Because the maximum
RM effect occurs when the planet orbit is aligned with the stellar rotation axis, we consider the misalignment angle to
be zero ($\lambda=0.$). We show the distances between the observer and the lens plane by $D_{\rm L}$,
between the observer and the source plane by $D_{\rm S}$, and between the lens plane and the source plane by $D_{\rm LS}$
(Figure 1). The radius of Einstein ring can be written as \citep{Sahu-03}

\begin{equation}
R_{E}=\sqrt{\frac{4GM_{L}D}{c^{2}}}, \,\,\,\,\,\,\,\,\,\,\,\,\, D=\frac{D_{LS}D_{L}}{D_{S}}\simeq D_{LS},
\end{equation}

\noindent
where $M_{L}$ is the mass of the lens (planet), $c$ is speed of light, and $G$ is the gravitational constant. As an example, the Einstein ring radius of a Jupiter-mass planet at 1 AU from a star
is  $\sim 0.0013 \, R_{\rm Sun}$. For an Earth-like planet at 1 AU, this value becomes $\sim 7.4 \times {10^{-5}}\,R_{\rm Sun}$.

In a system similar to that described above, the distance between a point on the stellar disk and the projection of the
center of the lens (planet) on the disk of the star can be written as

\begin{equation}
d=\sqrt{(x-x_{p})^{2}+(y-y_{p})^{2}}.
\end{equation}

\begin{figure}[h!]
  \centering
 \includegraphics[scale=0.7]{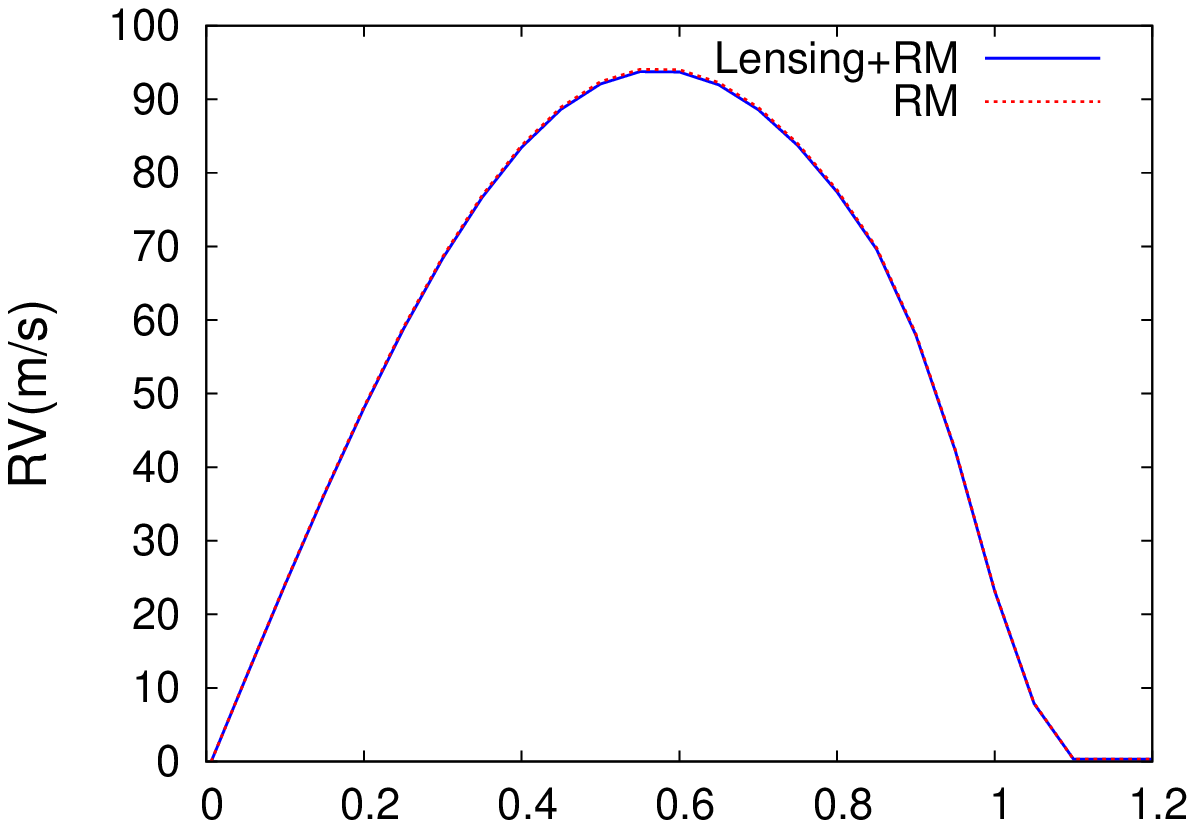}
  \centering
    \includegraphics[scale=0.7]{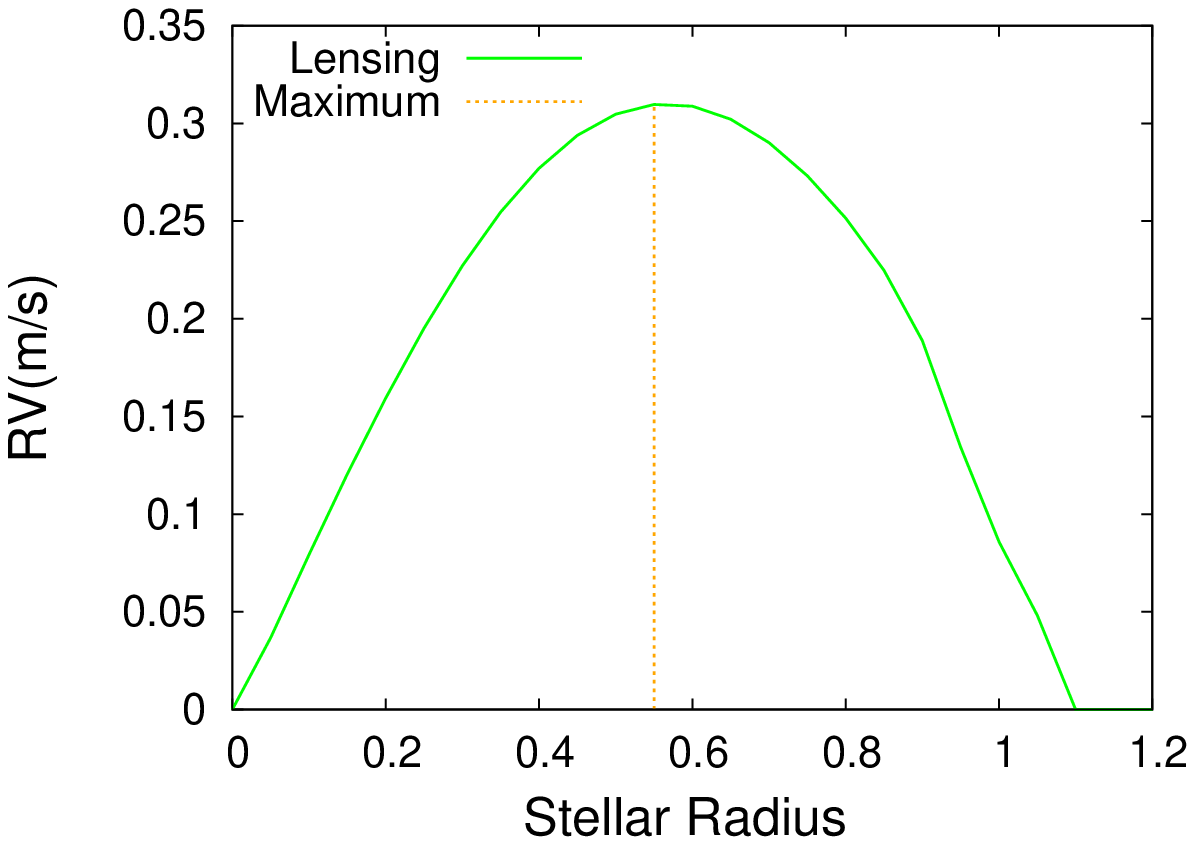}
  \caption{Top: Graphs of the RM effect with and without the contribution of the planet's microlensing. Note that the two
curves are almost overlaying each other. To remove the effect of microlensing, we assume $R_{E}=0$. Bottom: Graphs of residuals
between the two curves shown in the top plot. The residuals are indicative of the net behavior of microlensing as a function
of the planet's position along the stellar radius. The maximum value of the of microlensing appears at 0.55 stellar radii.}

\end{figure}

\noindent
The lens produces two images of the point $(x,y)$ along the line connecting this point to the center of the projection of the
planet on the star's disk. These images are at distances

\begin{equation}
d_{\pm}=0.5 \left[d\pm(d^{2}+4R_{E}^{2})^{1/2}\right]
\end{equation}

\noindent
from the point of the projection of the center of the lens (planet) on the stellar disk, and are visible when they are
not occulted by the planet (i.e., $|d_{\pm}|>r_{p}$) \citep{Sahu-03}. We, therefore, define a transmission function
$T(x,y)$ as

\begin{equation}
T(x_{\pm},y_{\pm}) =
\left\{
	\begin{array}{ll}
		0 & \mbox{if } |d_{\pm}| \leq r_{p} \\
		1   & \mbox{if } |d_{\pm}| > r_{p} \\
	\end{array}
\right.
\end{equation}

\noindent
to determine when the effect of the lensed-images have to be taken into account.
The amplification caused by these images is given by \citep{Sahu-03}

\begin{equation}
A_{\pm}(d)=\frac{1}{2}(1+2\,{\delta^{-2}}) \, \left[1+4\,{\delta^{-2}}\right]^{-1/2}\pm 0.5.
\end{equation}

\noindent
where $\delta=d/{R_E}$. In order to incorporate the effect of microlensing on the RV measurements (obtained from the RM effect)
the profile of the cross-correlation function (CCF) has to be calculated using the following integral
over the full stellar disk,

\begin{equation}
F(v)=\frac{\int \!\!\! \int K(x,y)M(x,y,v)dxdy}{\int \!\!\! \int K(x,y)dxdy}.
\end{equation}

\noindent
In this equation,

\begin{eqnarray}\nonumber
K(x,y) & = & \Big[A_{+}(x,y)\,T(x_{+},y_{+}) \left.
\right.\\
&& \left. + \, A_{-}(x,y)\, T(x_{-},y_{-})\Big] I(x,y)  \,\ , \right.
\end{eqnarray}

\noindent
is the modified stellar surface intensity, and

\begin{eqnarray}\nonumber
M(x,y,v)  =  \frac{1}{2\sqrt{\pi}} \!\!\! \!\! &&\left\{ \frac{1}{\sigma_{\parallel}}\,
{\rm Exp}\Biggl[{-\left(\frac{v-v(x,y)}{\sigma_{\parallel}}\right)^{2}}\Biggl]
\right.\\
&& \left. +\frac{1}{\sigma_{\perp}}\, {\rm Exp}\Biggl[{-\left(\frac{v-v(x,y)}{\sigma_{\perp}}\right)^{2}}\Biggl]  \right\} \,\ ,
\end{eqnarray}

\noindent
is the velocity profile associated with each point $(x,y)$ due to both the stellar rotation and
macro-turbulence \citep{Gray-05}. The quantities $\sigma_{\parallel}$ and $\sigma_{\perp}$ in equation (9) are given by
\begin{equation}
\sigma_{\parallel}^{2}=\sigma_{0}^{2}+ \zeta^{2}\cos\theta^{2} \,\,\,\,\,\, , \,\,\,\,\,\,
\sigma_{\perp}^{2}=\sigma_{0}^{2}+ \zeta^{2}\sin\theta^{2}\,,
\end{equation}

\noindent
where $\zeta$ is the macro-troublence velocity and is considered to have a value of $\zeta$=3.98 km\,${\rm s}^{-1}$,
equal to that of the Sun \citep{Valenti-05}. The quantity $\sigma_{0}$ is the instrumental broadening which is set to approximately
2.2 ~km\,${\rm s}^{-1}$. This is equivalent to the broadening of a typical spectrograph with a high-resolution
of $R \sim110000$ similar to that of HARPS. The velocity $v(x,y)$ in equation (9) is defined as

\begin{equation}
v(x,y)=(v\sin i)\frac{x}{r_{\ast}}+(v_{\rm CB})\mu\,,
\end{equation}

\noindent
where $v_{\rm CB}$ is the convective blue-shift and is set to -300 ~m\,${\rm s}^{-1}$, similar to its value for the Sun
\citep{Shporer-11}.

\section{Calculations and Results}

We considered a rotating star with $v\sin i=10 ~{\rm km\,s}^{-1}$ and a planet with a mass of $10 {M_{\rm Jup}}$.
We set the value of ${D_{\rm LS}}$ to 1 AU and let the position of the projection of the planet on the $(x,y)$ plane
vary from the outside of the disk of the star ($1.2 \, {r_{\ast}}$) to its center.

We used a gird of $4000\times 4000$ cells on the $(x,y)$ plane and calculated the integral of equation 7.
By fitting a Gaussian function to $F(v)$, we determined the center of CCF which corresponds to the value of RV
that takes into account the combined contributions from the RM and microlensing effects. To determine
the sole effect of microlensing on the value of RM, we carried out similar calculations considering the
Einstein ring radius to be zero ($R_{E}=0$). By subtracting the results from the previous calculations,
we determined the net contribution of the microlensing effect. Figure 2 shows the results. The bottom panel of this figure
shows the difference between the two calculations and the contribution of microlensing effect to the RV measurements.
As shown here, microlensing has its maximum contribution when the projection of the planet on the stellar disk
is at a distance of 0.55 stellar radius from the center of star. Because we are interested in the maximum effect of
microlensing, in the rest of this study, we hold the position of planet's projection at $(x_{\rm p},y_{\rm p})=(0.55 \, r_{\ast},0)$.

To better understand the effects of difference parameters on the results, we carried out calculations for different values of the mass and
orbital separation of the planet. The mass of the planet was taken to be 0.1, 0.5, 1, 10, and 50 $M_{\rm Jup}$ with its lowest
value corresponding to the mass of a Neptune-like object and its largest value being interpreted as the mass of a brown-dwarf.
The orbital separation ($D_{LS}$) was varied from 0.25 AU to 3 AU in steps of 0.25 AU. We considered the rotational velocity
of the host star to be $v\sin i=$ 2, 5, and 10 $~{\rm km \, s}^{-1}$ \citep{McNally-65}. We also considered three detection limits
for RV measurements: $ 50 \, {\rm cm\,s}^{-1}$ for current facilities such as HARPS, $10 \, {\rm \, cm \, s}^{-1}$ for observational
facilities in the near future such as ESPRESSO@VLT \citep{Pepe-10}, and $2 \, {\rm cm\,s}^{-1}$ for instruments farther in future
such as HIRES@E-ELT. As shown in figure 3, the microlensing amplification will have significant contribution to the RM
measurements of a massive planet in a long period orbit ($a > 2$ AU) when obtained from all the current and future observational
facilities. For instance,  in the case of a 10 Jupiter-mass planet in a 2 AU orbit around a star with $v\sin i=10 \, {\rm km\,s}^{-1}$,
the microlensing effect is about 1 ${\rm m\,s}^{-1}$ which is comparable to the effect of macro-turbulence and convective
blue-shift on RM measurements ($\sim 3-4 \, {\rm m\,s}^{-1}$) \citep{Albrecht-12}. In the case of a massive planet on relatively
short-period orbit ($<1$ AU), the results depend on the rate of the rotation of the star. If the star is fast rotating,
the microlensing effect will be significant, otherwise for slow rotating stars, this effect will be at the
detection limit of future facilities such as HIRES@E-ELT. We note that because the amplitude of the signal associated with the microlensing
effect on RM measurements for a planet with a mass of $0.1 M_{\rm Jup}$ and for all values of its orbital separations are smaller than
0.001 ${\rm m\,s}^{-1}$, we excluded these cases from the figure.

We also probed the impact of the microlensing on the measured values of the misalignment angle ($\lambda$),
and the stellar rotation velocity ($v\sin i$) obtained from the measurements of the RM effect. Our results indicated that
the misalignments angle will not be affected by microlensing. This is an expected results since the measurements of the
misalignment angle are sensitive to the asymmetry of the RM signal, and the microlensing effect does not
disturb the symmetry of this signal.

We would like to emphasize that in cases where the effect of microlensing is not negligible, both the light-curve
of the star and the RM signal have to be modeled taking this effect into account. In such systems, when fitting the light-curve,
ignoring the microlensing magnification will result in a planetary model in which the radius of the planet is smaller than its actual value.
Despite such a discrepancy in the modeled radius of the planet, the $v\sin i$
obtained from fitting the RM signal is only weakly affected. We note that in systems where the effect of
microlensing is included in fitting the light-curve \citep{Sahu-03}, this effect must also be included in the modeling of the
RM signal, otherwise the value of $v\sin i$ will be underestimated.

\section{Analytical Analysis}

Using the formalism presented by \citet{Ohta-05}, we have been able to derive an analytical formula that can
roughly reproduce the results of our simulations. As shown by these authors, the averaged value of the radial
velocity of the star is given by

\begin{equation}
v_{\rm av}=\frac{\int \!\!\! \int K(x,y)x \, dx \, dy} {\int \!\!\! \int K(x,y) \, dx \, dy}\, v\sin i.
\end{equation}

\noindent
To use equation (12) to determine the variation in the radial velocity (i.e., $\Delta$RV) due to the microlensing effect,
we recall that the microlensing effect depends only on the value of the Einstein radius $R_{E}$.
This quantity appears in the definition of $d_{\pm}$ and $A_{\pm}$. Taylor expanding these quantities
in terms of $\eta=R_{E}/d$, one can show that

\begin{equation}
d_{-} = \eta^{2}d \qquad , \qquad d_{+} = d+\eta^{2}d\,,
\end{equation}

\noindent
and

\begin{equation}
A_{-} = \eta^{4}  \qquad , \qquad A_{+} = 1+\eta^{4}\,
\end{equation}

\noindent
Equations (13) and (14) indicate that to the second order in $R_{E}$, $A_{-}=0$ and $A_{+} = 1$,
implying that to this order of approximation, the image at $d_{-}$ has no contribution to either the light-curve
or the RM effect. To the lowest order in $R_{E}$, the microlensing effect simply increases the fraction
of the stellar disk that is seen by the observer. In fact, for all points located behind the planet and at
distances between $r_{p}(1-\eta^{2})$ and $r_{p}$ from the center of the projection of the planet on the stellar
disk, $d_{+} > r_{p}$ implying that to the lowest order, the effect of microlensing is equivalent to reducing
the value of the planet radius from $r_{p}$ to $r_{p} - R_{E}^{2}/r_{p}$. Using equation 12, one then obtains

\begin{equation}
\Delta {\rm RV} \propto (R_{E}/r_{p})^{2} \, v\sin i\,.
\end{equation}

\noindent
The coefficient of proportionality in equation (15) can be determined using the results of our numerical
experiment. The change in the radial velocity is, therefore, given by

\begin{eqnarray}\nonumber
\Delta {\rm RV} ({\rm cm/s}) \approx 0.34\left(\frac{M_{L}}{M_{Jup}}\right)&&
\!\!\!\!\!\!\left(\frac{a}{\rm AU}\right)\\
&&\!\!\!\!\!\!\!\!\!\!\!\!\!\!\!\!\!\!
\left(\frac{R_{Jup}}{r_{p}}\right)^{2}\left(\frac{v\sin i}{\rm km/s}\right).
\end{eqnarray}

\section{Discussion and conclusion}

In this study, we explored and quantified the contribution of gravitational microlensing to the measurements of the radial
velocity during the transit of a planet (RM effect). We showed that the amplification of the RV signal due to the microlensing effect
depends strongly on the mass of the planet and its orbital separation. The results indicated
that the effect of microlensing is also significant for massive planets on long period orbits and is detectable with the current
observational facilities. Our results showed that microlensing does not disturb the measurements of the misalignments
angle caused by the RM effect. If when fitting the light-curve, this effect is taken into account, it
should also be considered in the modeling of the RM signal, otherwise the value of $v \sin i$ will be underestimated up to several
percent for extreme cases. We note that
in such cases, the underestimation of the rotational velocity will be within the observational error, and thus undetectable.

Most of the current RM surveys have been carried out on transiting systems that host planets in short period orbits.
To better constrain models of planet formation, studies of the RM effect have to also include massive planets in long periods.
As those planets orbit mostly fast-rotating stars [exoplanets.org], the contribution of microlensing to the measurements of
their RM effect will become considerable, and detectable with future observational facilities such as
ESPRESSO@VLT and HIRES@E-ELT.

\begin{figure*}[h!]
  \centering
    \includegraphics[scale=0.85]{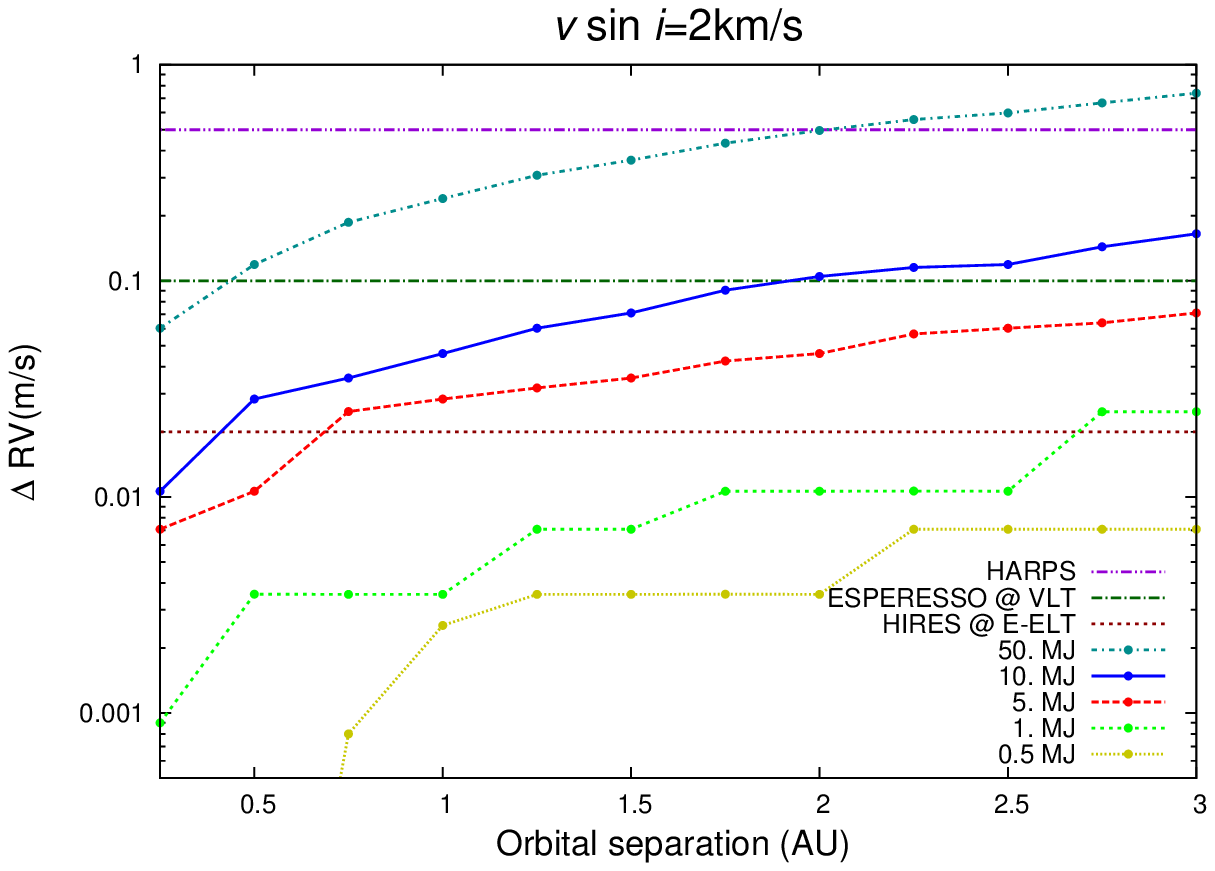}
  \centering
    \includegraphics[scale=0.85]{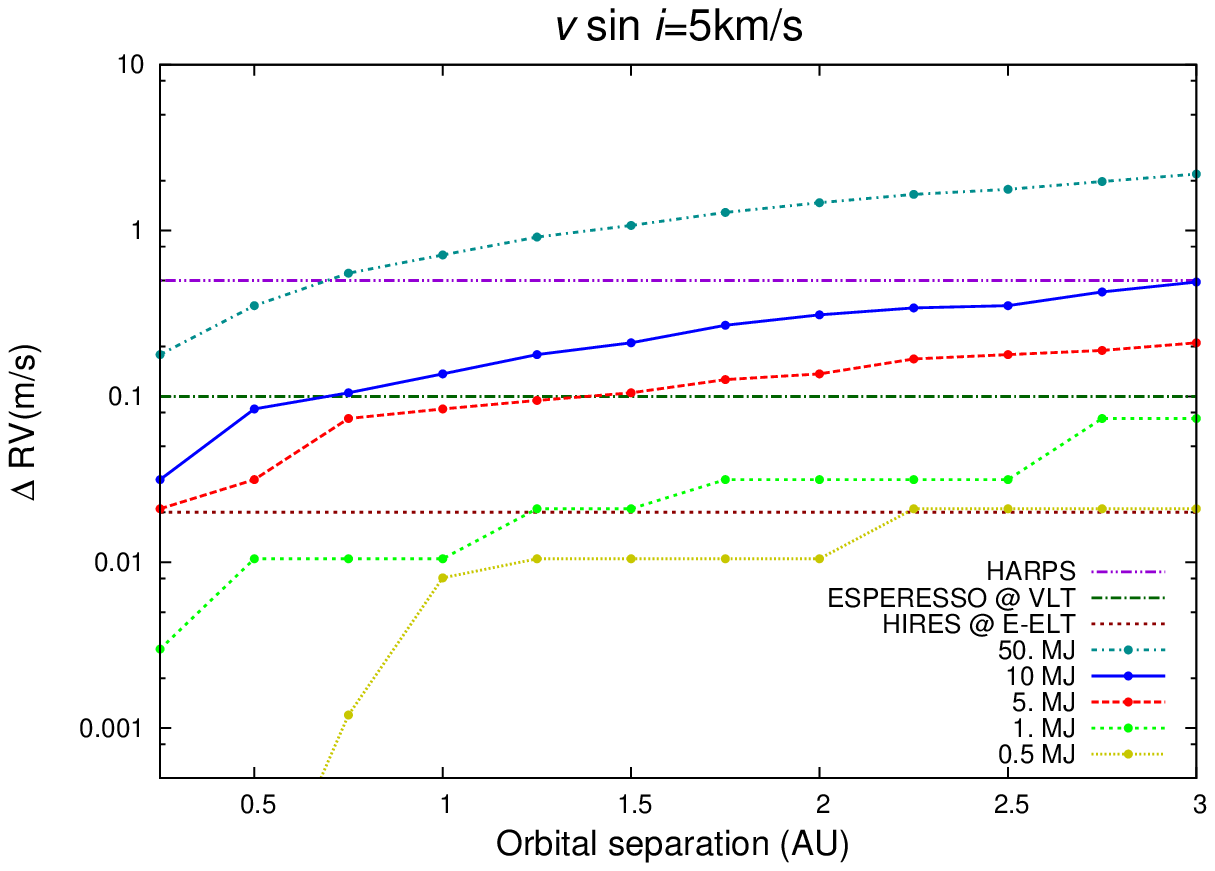}
  \centering
    \includegraphics[scale=0.85]{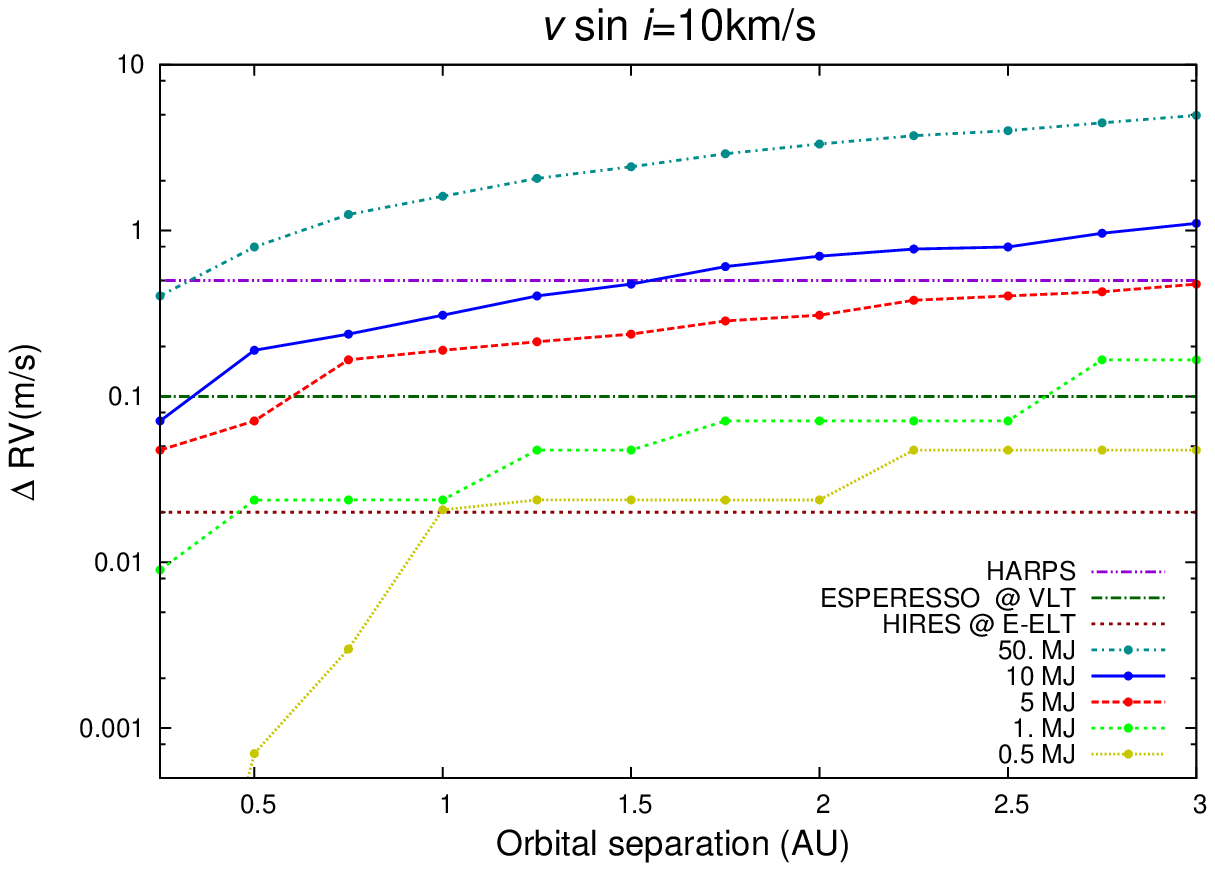}
  \caption{Maximum contribution of microlensing to the measurements of the RM effect for different combination of planet mass
and the position of the projection of the planet on the disk of the star. From top to bottom, the stellar rotation velocity
is equal to 2, 5, and 10$~{\rm km\,s}^{-1}$, respectively. The detection limit of HARPS, ESPRESSO @ VLT and
HIRES @ E-ELT are also shown for comparison.}
  \label{sample-figure}
\end{figure*}

\begin{acknowledgements}

We would like to thank R. Alonso for useful discussions. We acknowledge the support from the European Research Council/European Community under the
FP7 through Starting Grant agreement number 239953, and by Funda\c{c}\~ao para a Ci\^encia e a Tecnologia (FCT) in
the form of grants reference SFRH/BD/51981/2012. NCS also acknowledges
the support from FCT through program Ci\^encia\,2007 funded by FCT/MCTES (Portugal) and POPH/FSE (EC).
NH acknowledges support from the NAI under Cooperative Agreement NNA09DA77 at the Institute for Astronomy, University of
Hawaii, HST grant HST-GO-12548.06-A, and Alexander von Humboldt Foundation. NH is also thankful to the Institute for
Astronomy and Astrophysics/Department of Computational Physics at the University of Tuebingen, Germany for their
kind hospitality during the course of this project. Support for program
HST-GO-12548.06-A was provided by NASA through a grant from the Space
Telescope Science Institute, which is operated by the Association of Universities for Research in Astronomy, Incorporated,
under NASA contract NAS5-26555.
\end{acknowledgements}

\bibliographystyle{aa}
\bibliography{mahlibspot}

\begin{thebibliography}{19}
\expandafter\ifx\csname natexlab\endcsname\relax\def\natexlab#1{#1}\fi

\bibitem[{{Agol}(2003)}]{Agol-03}
{Agol}, E. 2003, \apj, 594, 449

\bibitem[{{Albrecht} {et~al.}(2012){Albrecht}, {Winn}, {Johnson}, {Howard},
  {Marcy}, {Butler}, {Arriagada}, {Crane}, {Shectman}, {Thompson}, {Hirano},
  {Bakos}, \& {Hartman}}]{Albrecht-12}
{Albrecht}, S., {Winn}, J.~N., {Johnson}, J.~A., {et~al.} 2012, \apj, 757, 18

\bibitem[{{Brown} {et~al.}(2001){Brown}, {Charbonneau}, {Gilliland}, {Noyes},
  \& {Burrows}}]{Brown-01}
{Brown}, T.~M., {Charbonneau}, D., {Gilliland}, R.~L., {Noyes}, R.~W., \&
  {Burrows}, A. 2001, \apj, 552, 699

\bibitem[{{Einstein}(1936)}]{Einstein-36}
{Einstein}, A. 1936, Science, 84, 506

\bibitem[{{Gray}(2005)}]{Gray-05}
{Gray}, D.~F. 2005, {The Observation and Analysis of Stellar Photospheres}

\bibitem[{{H{\'e}brard} {et~al.}(2008){H{\'e}brard}, {Bouchy}, {Pont},
  {Loeillet}, {Rabus}, {Bonfils}, {Moutou}, {Boisse}, {Delfosse}, {Desort},
  {Eggenberger}, {Ehrenreich}, {Forveille}, {Lagrange}, {Lovis}, {Mayor},
  {Pepe}, {Perrier}, {Queloz}, {Santos}, {S{\'e}gransan}, {Udry}, \&
  {Vidal-Madjar}}]{Hebrard-08}
{H{\'e}brard}, G., {Bouchy}, F., {Pont}, F., {et~al.} 2008, \aap, 488, 763

\bibitem[{{Hirano} {et~al.}(2011){Hirano}, {Suto}, {Winn}, {Taruya}, {Narita},
  {Albrecht}, \& {Sato}}]{Hirano-11}
{Hirano}, T., {Suto}, Y., {Winn}, J.~N., {et~al.} 2011, \apj, 742, 69

\bibitem[{{McLaughlin}(1924)}]{McLaughlin-24}
{McLaughlin}, D.~B. 1924, \apj, 60, 22

\bibitem[{{McNally}(1965)}]{McNally-65}
{McNally}, D. 1965, The Observatory, 85, 166

\bibitem[{{Muirhead} {et~al.}(2013){Muirhead}, {Vanderburg}, {Shporer},
  {Becker}, {Swift}, {Lloyd}, {Fuller}, {Zhao}, {Hinkley}, {Pineda}, {Bottom},
  {Howard}, {von Braun}, {Boyajian}, {Law}, {Baranec}, {Riddle}, {Ramaprakash},
  {Tendulkar}, {Bui}, {Burse}, {Chordia}, {Das}, {Dekany}, {Punnadi}, \&
  {Johnson}}]{Muirhead-13}
{Muirhead}, P.~S., {Vanderburg}, A., {Shporer}, A., {et~al.} 2013, \apj, 767,
  111

\bibitem[{{Ohta} {et~al.}(2005){Ohta}, {Taruya}, \& {Suto}}]{Ohta-05}
{Ohta}, Y., {Taruya}, A., \& {Suto}, Y. 2005, \apj, 622, 1118

\bibitem[{{Pepe} {et~al.}(2010){Pepe}, {Cristiani}, {Rebolo Lopez}, {Santos},
  {Amorim}, {Avila}, {Benz}, {Bonifacio}, {Cabral}, {Carvas}, {Cirami},
  {Coelho}, {Comari}, {Coretti}, {de Caprio}, {Dekker}, {Delabre}, {di
  Marcantonio}, {D'Odorico}, {Fleury}, {Garc{\'{\i}}a}, {Herreros Linares},
  {Hughes}, {Iwert}, {Lima}, {Lizon}, {Lo Curto}, {Lovis}, {Manescau},
  {Martins}, {M{\'e}gevand}, {Moitinho}, {Molaro}, {Monteiro}, {Monteiro},
  {Pasquini}, {Mordasini}, {Queloz}, {Rasilla}, {Rebord{\~a}o}, {Santana
  Tschudi}, {Santin}, {Sosnowska}, {Span{\`o}}, {Tenegi}, {Udry}, {Vanzella},
  {Viel}, {Zapatero Osorio}, \& {Zerbi}}]{Pepe-10}
{Pepe}, F.~A., {Cristiani}, S., {Rebolo Lopez}, R., {et~al.} 2010, in Society
  of Photo-Optical Instrumentation Engineers (SPIE) Conference Series, Vol.
  7735, Society of Photo-Optical Instrumentation Engineers (SPIE) Conference
  Series

\bibitem[{{Rossiter}(1924)}]{Rossiter-24}
{Rossiter}, R.~A. 1924, \apj, 60, 15

\bibitem[{{Sahu} \& {Gilliland}(2003)}]{Sahu-03}
{Sahu}, K.~C. \& {Gilliland}, R.~L. 2003, \apj, 584, 1042

\bibitem[{{Shporer} \& {Brown}(2011)}]{Shporer-11}
{Shporer}, A. \& {Brown}, T. 2011, \apj, 733, 30

\bibitem[{{Simpson} {et~al.}(2010){Simpson}, {Pollacco}, {H{\'e}brard},
  {Gibson}, {Barros}, {Boisse}, {Bouchy}, {Collier Cameron}, {Miller},
  {Watson}, \& {Keenan}}]{Simpson-10}
{Simpson}, E.~K., {Pollacco}, D., {H{\'e}brard}, G., {et~al.} 2010, \mnras,
  405, 1867

\bibitem[{{Valenti} \& {Fischer}(2005)}]{Valenti-05}
{Valenti}, J.~A. \& {Fischer}, D.~A. 2005, \apjs, 159, 141

\bibitem[{{Winn} {et~al.}(2009){Winn}, {Johnson}, {Fabrycky}, {Howard},
  {Marcy}, {Narita}, {Crossfield}, {Suto}, {Turner}, {Esquerdo}, \&
  {Holman}}]{Winn-09}
{Winn}, J.~N., {Johnson}, J.~A., {Fabrycky}, D., {et~al.} 2009, \apj, 700, 302

\bibitem[{{Winn} {et~al.}(2010){Winn}, {Johnson}, {Howard}, {Marcy},
  {Isaacson}, {Shporer}, {Bakos}, {Hartman}, \& {Albrecht}}]{Winn-10}
{Winn}, J.~N., {Johnson}, J.~A., {Howard}, A.~W., {et~al.} 2010, \apjl, 723,
  L223

\end{thebibliography}

\end{document}